\documentclass[]{raa}

\usepackage{graphicx,times}
\usepackage{natbib}
\usepackage{amssymb,amsmath}
\bibpunct{(}{)}{;}{a}{}{,}

\usepackage{threeparttable}

\usepackage{CJKutf8}

\def\s{\,{\rm s}}

\def\erg{\,{\rm erg}}
\def\kev{\,{\rm keV}}

\usepackage[colorlinks,citecolor=cyan]{hyperref}

\begin{document}

  \title{Joint constraint on the jet structures from the short GRB population and GRB 170817A}

   \volnopage{Vol.0 (20xx) No.0, 000--000}
   \setcounter{page}{1}

   \author{Xiao-Feng Cao 
      \inst{1}
   \and Wei-Wei Tan 
      \inst{1}
   \and Yun-Wei Yu 
      \inst{2, 3, 4}
    \and Zhen-Dong Zhang 
      \inst{5} 
      }

   \institute{Research Center for Astronomy, Hubei University of Education, Wuhan 430205, China; {\it  caoxf@mails.ccnu.edu.cn; wwtan@hue.edu.cn}\\
        \and
             Institute of Astrophysics, Central China Normal University, Wuhan 430079, China; {\it yuyw@ccnu.edu.cn}\\
        \and
             Key Laboratory of Quark and Lepton Physics (Central China Normal University), Ministry of Education, Wuhan 430079, China\\
        \and 
             Education Research and Application Center, National Astronomical Data Center, Wuhan 430079, China\\
        \and 
            Department of Physics and Astronomy, University of Tennessee, Knoxville, TN 37996-1200, USA\\
\vs\no
   {\small Received 2025 month day; accepted 2025 month day}}

\abstract{The nearest GRB 170817A provided an opportunity to probe the angular structure of the jet of this short gamma-ray burst (SGRB), by using its off-axis observed afterglow emission. It is investigated that whether the afterglow-constrained jet structures can be consistent with the luminosity of the prompt emission of GRB 170817A. Furthermore, by assuming that all SGRBs including GRB 170817A have the same explosive mechanism and jet structure, we apply the different jet structures to the calculation of the flux and redshift distributions of the SGRB population, confronting with the observational distributions of the Swift and Fermi sources. In comparison, it is found that the power-law and double-Gaussian models could be somewhat more favored than the single-Gaussian structure, as the last one could be not very consistent with the flux distribution of the Fermi data. Nevertheless, in view of the large uncertainty of parameters, we are still unable to make a sufficiently conclusive judgment at present.  
}

   \authorrunning{Cao et al. 2025 }
   \titlerunning{Joint constraint on SGRB jet structures}

   \maketitle

\section{Introduction}
Gamma-ray bursts (GRBs) are generated by highly-beamed relativistic jets, which are driven by rapidly rotating black hole or neutron star engines. Before gamma-ray emission is produced, the jets should first propagate through dense progenitor material, which can be a stellar envelope for long GRBs \citep{Zhang2003,Matzner2003,Lazzati2005,Bromberg2011,Bromberg2014,Suwa2011,Yu2020,Gottlieb2022,Urrutia2023a,Urrutia2023b} or a merger ejecta for short GRBs \citep[SGRBs;][]{Nagakura2014,Lazzati2017,Yu2020,Hamidani2021,Nathanail2021,Nativi2022,Gottlieb2022,Pavan2023}.

As a result of its interaction with the progenitor material, a GRB jet breaking out from the progenitor can finally own an angular structure for its energy and velocity distributions. Generally speaking, the breakout jet can consist of a core region with an opening angle of few degrees and a relatively wider and less energetic wing region \citep{Lazzati2018,Salafia2022}. In addition, the jet can also be surrounded by a much wider cocoon that is contributed by the shocked material. Nevertheless, it seems unnecessary to treat the jet and cocoon separately, due to the mixing of the material and the continuous distribution of the energy. Instead, we can simply treat the cocoon as a part of the jet wing. From the core to the wing, the Lorentz factor and the energy density of the jet can decrease rapidly with the increasing angle relative to the jet axis. Three empirical analytical functions have usually been suggested to describe the angular structure of GRB jets, including a power law \citep{Dai2001,Zhang2002,Kumar2003,Lazzati2005}, a Gaussian \citep{Zhang2002,Rossi2002,Rossi2004, Kumar2003,Granot2003} and, sometimes, double Gaussians \citep{Tan2020,Luo2022,Wei2022}.

Observational constraints on the jet structures can in principle provide a clue to understand the nature and interior of GRB progenitors. First, a direct implication for the angular structure can be derived from the afterglow emission of GRBs, particularly, when the viewing direction deviates significantly from the jet axis \citep{Kumar2003}. Because of the relativistic beaming effect, the emission from the jet material deviating from the line of sight (LOS) can be detected only after the material is decelerated to have an emission beaming angle larger than the viewing angle. Therefore, it can be expected that the more luminous emission from more energetic jet material can be detected later for an off-axis observation. In this case, the peak of the afterglow light curves can appear when the core emission comes into sight and the increasing light curves before the peak just reflect the angular distribution of the jet energy density.

The problem is that, for the majority of observed GRBs, their cosmological distances always prevent us from detecting them on a large viewing angle, because of the rapid decrease of the jet energy with the angle. And, without a GRB trigger, it will be very difficult to capture the orphan afterglow emission of the GRBs. Following such a consideration, the observed GRBs are always assumed to be on-axis and a ``top-hat" structure is generally appropriate for the afterglow modelings, at most, by further invoking an opening angle for the jet if a so-called jet-break feature appears in the light curves.
Nevertheless, this situation has being changed since the detection of the nearest GRB 170817A of a distance of $\sim40$ Mpc. The viewing angle of this GRB was quickly constrained to be about $\theta_{\rm obs}\leq31^{\circ}$ by the gravitational wave detection of GW 170817 \citep{Abbott2017a}. This special multi-messenger event provided the first opportunity to constrain the angular structure of the GRB jet by its afterglow emission and various jet structure models had been widely investigated \citep{Lamb2017, Xiao2017, Margutti2017, Margutti2018, Troja2017, Troja2018, D'Avanzo2018, Lazzati2018,Mooley2018,Granot2018, Resmi2018, Xie2018, Nynka2018, He2018, Ziaeepour2019, Huang2019, Kathirgamaraju2019, Lamb2019,Beniamini2019, Beniamini2020,  Takahashi2020, Takahashi2021, Wei2022}. Besides explaining the afterglow emission, the off-axis observation of GRB 170817A also provides a natural but qualitative explanation for its ultra-low luminosity of $\sim10^{47}\rm erg~s^{-1}$ \citep{Abbott2017b,Zhang2018}. Then, it needs to be checked whether the observed prompt luminosity can be quantitatively consistent with the jet structure derived from the afterglow modelings.

Furthermore, a large viewing angle of GRB 170817A is necessary to understand the very high event rate of nearby GRBs inferred from GRB 170817A. Meanwhile, a question arises here: How can we connect this nearby SGRB rate in a natural way with the rates of the other SGRBs? In more detail, if all SGRBs including GRB 170817A share a common geometry for their jets, then it is crucial to ask how this common jet structure influences our understanding of the observational redshift and energy distributions of all SGRBs as well as the determination of the luminosity function (LF) and event rate of the SGRBs, as previously investigated \citep{Salafia2020,Salafia2022,Tan2020,Luo2022,Hayes2023}. It should be noted that the luminosity can actually vary with the angle even in the jet core region, but is not uniformly distributed as assumed in the top-hat model. Meanwhile, the LOS of SGRBs is random distributed and usually not strictly parallel to the jet axis. Therefore, even though some SGRBs have an identical jet, their observed luminosity can still be different and distributed within a range. In this case, if a top-hat uniform jet structure is preconceived in our mind, then we would ascribe this luminosity distribution to be a component of the LF of SGRBs. This is the reason why an apparent broken power law LF can always be obtained. In contrast, if a realistic jet structure can be taken into account correctly, then we can find that the low-luminosity component of the apparent LF can naturally result from the random distribution of the viewing angles \citep{Tan2020}. Meanwhile, it is completely sufficient to invoke the high-luminosity power law to describe the distribution of the central luminosity of SGRB jets. In other words, the intrinsic LF of SGRBs can be a single-power law.

\begin{figure*}
	\centering
	\includegraphics[width=0.33\textwidth]{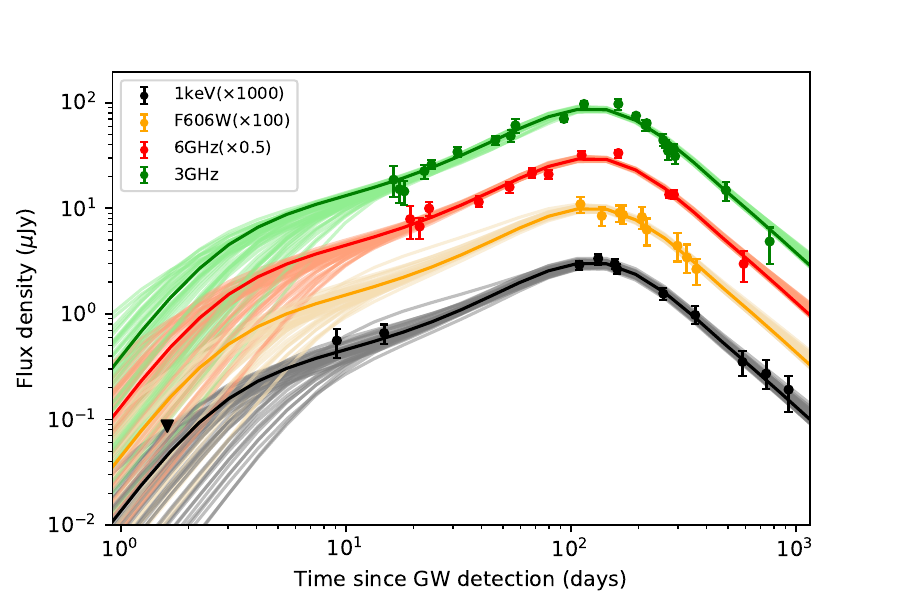}\includegraphics[width=0.33\textwidth]{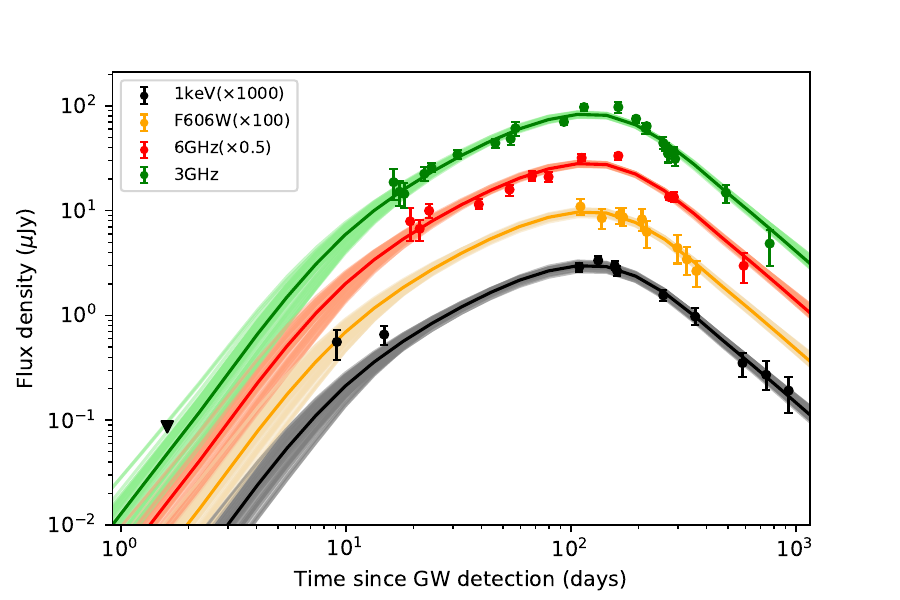}\includegraphics[width=0.33\textwidth]{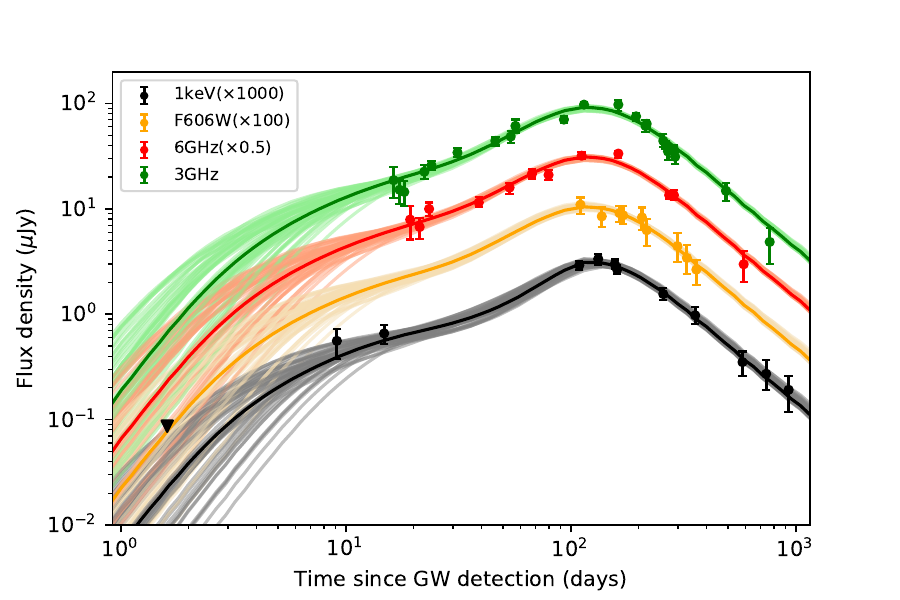}\\
	\caption{The fits of the multi-wavelength afterglow light curves of GRB 170817A for a power-law (left), single-Gaussian (middle), or double-Gaussian (right) jet structure. The observational data are taken from \citet{Makhathini2021}.}
	\label{fig1}
\end{figure*}

\begin{table*}
	\centering
	\setlength{\tabcolsep}{7mm}{}
	\renewcommand\arraystretch{1.4}
	\caption{Model parameters constrained from the afterglow modeling of GRB170817A. }\label{tab1}
	\begin{tabular}{lllll}
		\hline \hline
		Parameter &   Prior &&Posterior &\\ 
        &&Power-law  &Single-Gaussian        &Double-Gaussian\\
		\hline
		$\theta_{\rm obs} /\rm deg$   & (17, 35)    & $22.89_{-3.03}^{+3.71}$   & $23.21_{-3.60}^{+3.95} $ & $19.37_{-1.53}^{+2.21} $  \\
		$\log(\varepsilon_{\rm c}/\rm{erg~sr^{-1}})$      & (48, 53)   & $50.92_{-0.54}^{+0.35}$   & $51.76_{-0.98}^{+0.85}$  & $51.04_{-0.31}^{+0.24}$  \\
		$\log(C_{\rm E})$             & (-5, 0)       & $/$   & $/$                      & $-1.33_{-0.18}^{+0.19}$   \\
		$\Gamma_{\rm c}$      & (100, 600)      & $352_{-174}^{+265}$   & $507_{-103}^{+66} $      & $456_{-235}^{+236} $         \\
		$C_{\Gamma}$       & (0, 1)                 & $/$   & $/$                      & $0.49_{-0.30}^{+0.33}$     \\
		$\theta_{\rm{c}} /\rm deg$       & (1, 10)  & $2.56_{-0.62}^{+1.08}$   & $3.64_{-0.53}^{+0.63}$   & $1.48_{-0.32}^{+0.41}$ \\
		$\theta_{\rm{out}}/\rm deg$   & (2, 15)   & $/$   & $/$                      & $4.16_{-0.54}^{+0.87} $  \\
		$k_1$   & (3, 8)   & $3.97_{-0.52}^{+0.64}$  & $/$                      & $/$    \\
		$k_2$   & (0.1, 3) & $2.64_{-0.42}^{+0.25}$   & $/$                      & $/$     \\
		$\log (n/\rm cm^{-3})$     & (-4, 0)       & $-2.98_{-0.57}^{+0.49}$   & $-2.06_{-1.01}^{+0.94} $ & $-3.19_{-0.43}^{+0.40} $   \\
		
		$\log(\epsilon_{\rm B})$       & (-6, -1)       & $-2.36_{-0.46}^{+0.63}$   & $-3.98_{-1.11}^{+1.31} $ & $-2.59_{-0.28}^{+0.41} $  \\
		$p$                         & (2, 2.3)       & $2.13_{-0.01}^{+0.01}$   & $2.12_{-0.01}^{+0.01} $  & $2.13_{-0.01}^{+0.01} $   \\
		\hline	
	\end{tabular}\\
\end{table*}

However, in the previous work, the constraints on the jet structure from the GRB 170817A observations and population statistics were usually treated separately but have not been quantitatively confronted with each other. Therefore, this paper is devoted to investigating the consistency between these two types of observational constraints. In the next section, we briefly introduce the afterglow model and display the constrained model parameters for three typical structure functions. In Section 3, on the one hand, we derive the angular dependence of the equivalent isotropic emission energy ($E_{\gamma,\rm iso}$) from the energy density ($\varepsilon$) distribution of the jets, compared to the prompt luminosity of GRB 170817A. On the other hand, we compare the model-predicted redshift and flux distributions of SGRBs with the observational ones. A summary is given in Section 4.

\section{Constraining jet structure by the afterglows of GRB 170817A}
As usual, for calculating the afterglow emission from a structured jet, we can separate the jet into a series of differential rings and the dynamical evolution of each ring can be expressed as \citep{Huang1999,Li2019}:
\begin{equation}
	\frac{d\Gamma_{\theta}}{dM_{\rm sw,\theta}}=-\frac{\Gamma_{\theta}^{2}-1}{M_{\rm ej,\theta}+2\Gamma_{\theta} M_{\rm sw,\theta}},\label{EQ: dyn}
\end{equation}
where $\Gamma_{\theta}$ is the Lorentz factor of the ring of an half-opening angle $\theta$, $M_{\rm ej,\theta}$ and $M_{\rm sw,\theta}$ are the masses per solid angle of the GRB ejecta and the swept-up interstellar medium (ISM), respectively. Here, the possible lateral motion of the jet material is ignored, and thus the dynamical evolution of each ring can be considered independently, which had also been widely considered in previous afterglow modelings of GRB 170817A \citep[e.g.,][]{Lazzati2018,Gill2018,Morsony2024}. It is sometimes suggested that a top-hat jet could expand laterally on the sound speed of the shocked material \cite[e.g.,][]{Huang1999}. However, numerical simulations show that this could not be true and, instead, most of the jet energy can actually remain within the initial opening angle of the jet for a long period \citep{Zhang2009,Meliani2010,vanEerten2010,Granot2012}. By denoting the angular distribution of the jet kinetic energy by $\varepsilon_{\theta}\equiv dE_{\rm k}/d\Omega$, we have $M_{\rm ej,\theta}=\varepsilon_{\theta}/\Gamma_{\theta,\rm i}c^2$, where the subscript ``i" of $\Gamma_{\theta,\rm i}$ represents its initial value. Meanwhile, the increase of the swept-up mass is determined by
\begin{equation}
	\frac{dM_{\rm sw,\theta}}{dr_{\theta}}=r_{\theta}^{2}n m_{\rm p},
\end{equation}
where $r_{\theta}$ is the radius of the jet external shock, $n$ is the number density of the ISM which is considered to be constant in our calculation, and $m_{\rm p}$ is the mass of proton.

Following \cite{Sari1998}, the synchrotron luminosity contributed by a differential element of a mass $M_{\rm sw,\theta}$ can be calculated analytically as \citep{Yu2022}
\begin{equation}
	I'_{\nu'}(r,\theta,\varphi)={M_{\rm sw,\theta}\over r_{\theta}^2m_{\rm p}}{m_{\rm e}c^2\sigma_{\rm T}B'\over 12\pi e}S(\nu'),
\end{equation}
where the superscript prime indicates the quantities are measured in the comoving frame of the shocked region, $m_{\rm e}$ is the mass of electron, $c$ the speed of light, $\sigma_{\rm T}$ the Thomson cross section, $e$ the electron charge, and $B'$ represents the comoving magnetic field strength. The dimensionless synchrotron spectrum $S(\nu')$ can be expressed as a broken power law function, which is characterized by two broken frequencies that are determined by the acceleration and cooling of electrons (see \cite{Sari1998} for details). As usual, three microphysical parameters are involved here, including the spectral index of the shock-accelerated electrons $p$, the equipartition factors $\epsilon_{\rm B}$ and $\epsilon_{\rm e}$ for the magnetic fields and electrons in the shocked material, respectively. Moreover, a relationship $\epsilon_{\rm e}=\sqrt{\epsilon_{\rm B}}$ will be taken in our calculation in order to reduce the degree of freedom of the model, according to the theoretical prediction given by \cite{Medvedev2006}. Then, for an observer of a viewing angle $\theta_{\rm obs}$ relative to the jet axis, the observed flux of the afterglow emission can be obtained by integrating over the whole solid angle of the jet as \citep{Huang2000,Yu2007,Yu2022}
\begin{equation}
	F_{\rm \nu}(t)={r_{\theta}^2\over d^{2}_{\rm L}}\int \frac{I^{'}_{\nu^{'}}(r,\theta,\phi)}{\Gamma_{\theta}^3(1-\beta_{\theta}\cos\alpha)^3}\cos\alpha d\Omega(\theta,\phi),
\end{equation}
where $d_{\rm L}$ is the luminosity distance of the GRB, $\beta_{\theta}=(1-\Gamma_{\theta}^{-2})^{1/2}$, and $\alpha$ is defined as the angle between the emitting differential element and the LOS, which can be expressed as \citep[e.g.,][]{Li2019}
\begin{eqnarray}
	\cos\alpha  =\cos\theta\cos\theta_{\rm obs}+\sin\theta\sin\theta_{\rm obs}\cos\phi.
\end{eqnarray}
Finally, the connection between the radius of emitting material and the observational time is given by ${dr_{\theta}/ dt}={\beta_{\theta}c/( 1-\beta_{\theta}\cos\alpha)}$.

In this paper, the following three representative functions are taken to describe the possible jet structures:
\begin{enumerate}
	\item  Power-law jet
	\begin{eqnarray}
		\varepsilon_\theta&=&\varepsilon_{\rm c}\Theta^{-k_1},\\
		\Gamma_{\theta,\rm i}&=&\Gamma_{\rm c}\Theta^{-k_2}+1,
	\end{eqnarray}
	with $\Theta=\left[1+({\theta_{\rm }/ \theta_{\rm c}})^2\right]^{1/2}$;
	\item  Single-Gaussian jet
	\begin{eqnarray}
		\varepsilon_\theta&=&\varepsilon_{\rm c}\exp\left(-{\theta_{\rm }^2\over 2\theta_{\rm c}^2}\right),\\
		\Gamma_{\theta,\rm i}&=&\Gamma_{\rm c}\exp\left(-{\theta_{\rm }^2\over 2\theta_{\rm c}^2}\right)+1;
	\end{eqnarray}
	\item  Double-Gaussian jet
	\begin{eqnarray}
		\varepsilon_\theta&=&\varepsilon_{\rm c}\left[\exp\left(-{\theta_{\rm }^2\over 2\theta_{\rm c}^2}\right)+\mathcal C_{\rm E}\exp\left(-{\theta_{\rm }^2\over 2\theta_{\rm out}^2}\right)\right],\\
		\Gamma_{\theta,\rm i}&=&\Gamma_{\rm c}\left[\exp\left(-{\theta_{\rm }^2\over 2\theta_{\rm c}^2}\right)+\mathcal C_{\Gamma}\exp\left(-{\theta_{\rm }^2\over 2\theta_{\rm out}^2}\right)\right]+1,\nonumber\\
  &&
	\end{eqnarray}
\end{enumerate}
where the free parameters {\bf $k_1$, $k_2$,} $\varepsilon_{\rm c}$, $\Gamma_{\rm c}$, $\theta_{\rm c}$, $C_{\rm E}$, $C_{\Gamma}$, and $\theta_{\rm out}$ can be constrained by fitting the observed afterglows of GRB 170817A. Substituting these structure functions into the dynamical equation, we calculate the afterglow light curves for arbitrary viewing angles and constrain the model parameters by confronting the observational data of GRB 170817A, as presented in Figure \ref{fig1} for a direct impression\footnote{The result corresponding to the double-Gaussian case has been previously presented in \cite{Wei2022}.}. Here, since the flux evolution for an off-axis observation basically traces the angular dependence of the jet properties, it is in principle possible to obtain a constraint on the structure parameters, even in the double-Gaussian case where the outer and inner Gaussian components can correspond to the early and late light curves roughly separately.

To be specific, the constraints on the model parameters are obtained with Bayesian method, where the \textit{emcee} package \citep{Foreman-Mackey2013} is used within a wide range of parameter values.
The priors and posteriors with $1\sigma$ errors of the parameter values are listed in Table \ref{tab1}, which are in good agreement with the previous results \citep[e.g.,][]{Lazzati2018,Gill2018,Morsony2024}. Here, the log-likelihood function is defined as \( \ln \mathcal{L} = -\frac{1}{2} \sum_{i=1}^{n} \frac{(O_i - M_i)^2}{\sigma_i^2} \), where \( O_i \) are the observations, \( M_i \) are the corresponding model predictions, and \( \sigma_i \) are the observational uncertainties. To ensure comprehensive mixing and robust convergence of the MCMC sampling, the chains were allowed to run for more than 50 times of the autocorrelation time, where the autocorrelation times were found to be approximately 100 for each parameter. Furthermore, samplings were repeated multiple times to ensure consistent convergence, leading to a reliable and reproducible solution.  In view of the large number of model parameters (that is, 9, 7, and 10 for the power-law, single-Gaussian, and double-Gaussian cases, respectively), the parameter constraints are always subject to a degeneracy problem \citep[e.g.,][]{Granot2002,Nakar2021}, which cannot be eliminated even though the \textit{emcee} package is used \citep{Garcia-Cifuentes2024}. Then, undoubtedly, it is necessary to find more independent constraints. For example, \cite{Fong2019} measured $ p =2.166\pm 0.026$ directly from the evolving afterglow spectra from radio to X-ray, which is not very different from our result of $p\sim (2.12-2.13)$. Moreover, the dynamical evolution of the structured jet could also be inferred from the superluminal apparent motion of the centroid of radio emission sampling detected by the very long-baseline interferometry (VLBI) \citep{Mooley2018}. Therefore, in the next section, we will further endeavor to constrain these jet structures according to the statistical distributions of SGRBs.

\section{Confronting the jet structures with the population statistics of SGRBs}
\subsection{The direction-dependence of the emission energy}
\begin{figure}
	\centering
	\includegraphics[width=0.6\textwidth]{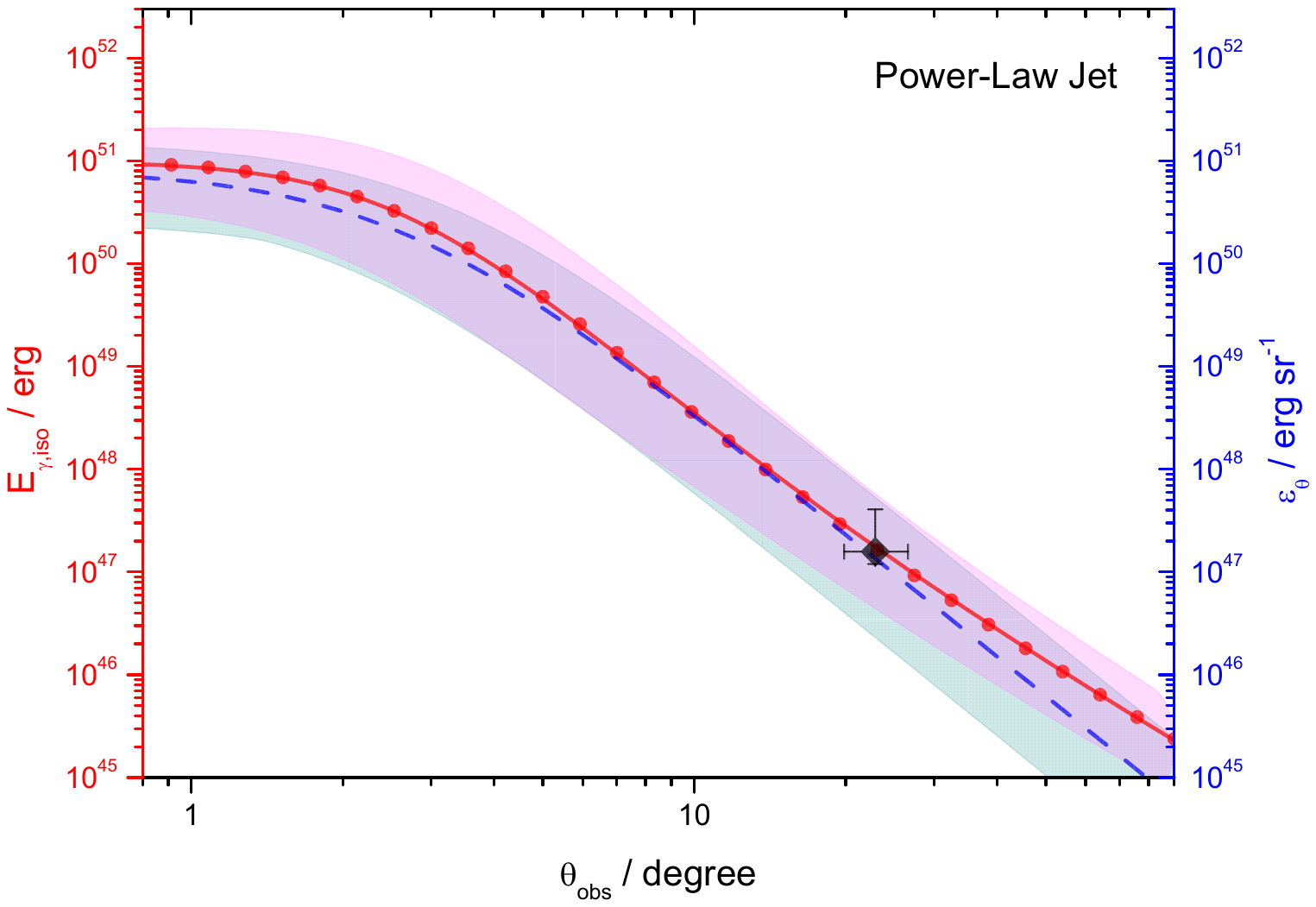}\\
 \includegraphics[width=0.6\textwidth]{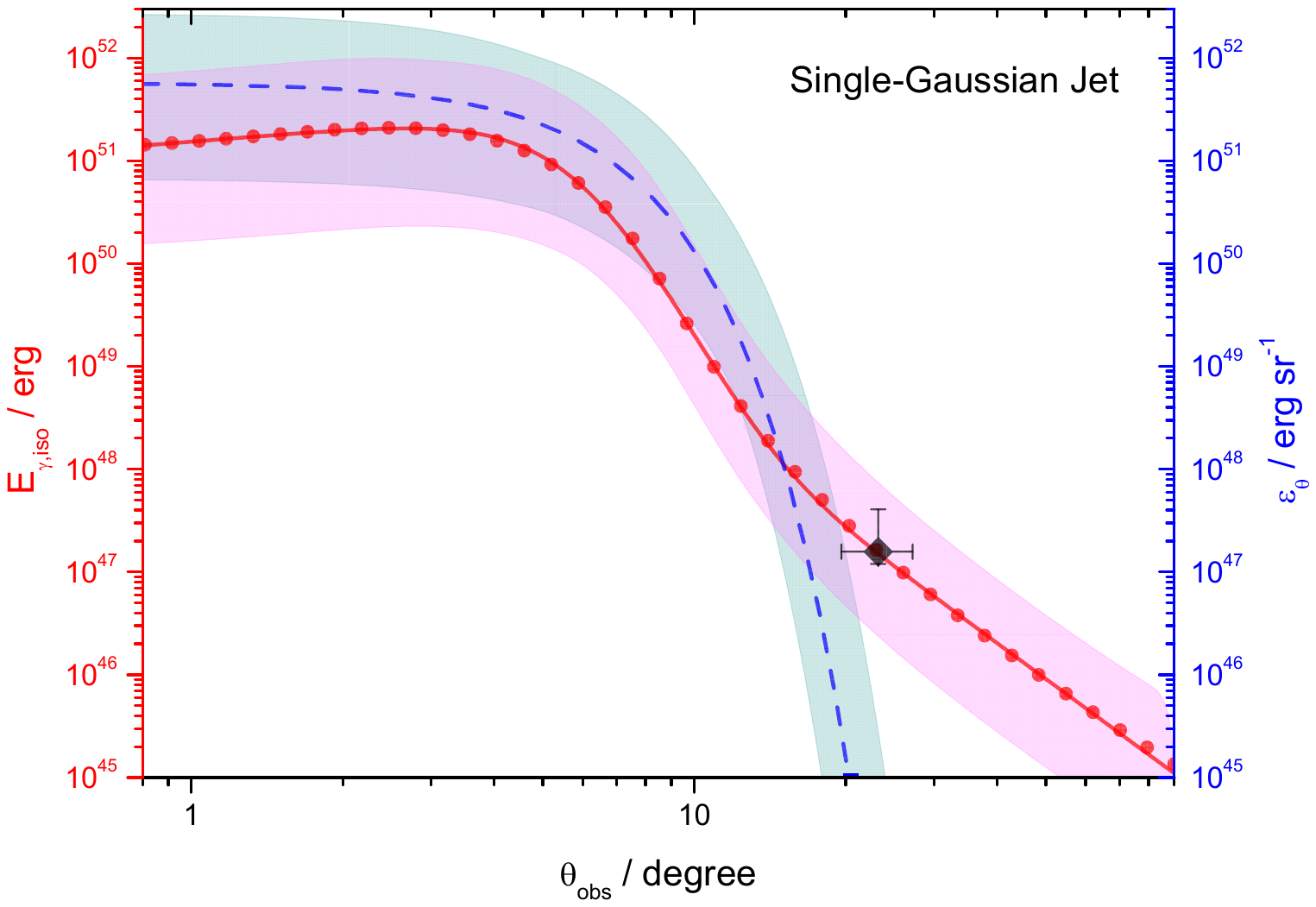}\\
 \includegraphics[width=0.6\textwidth]{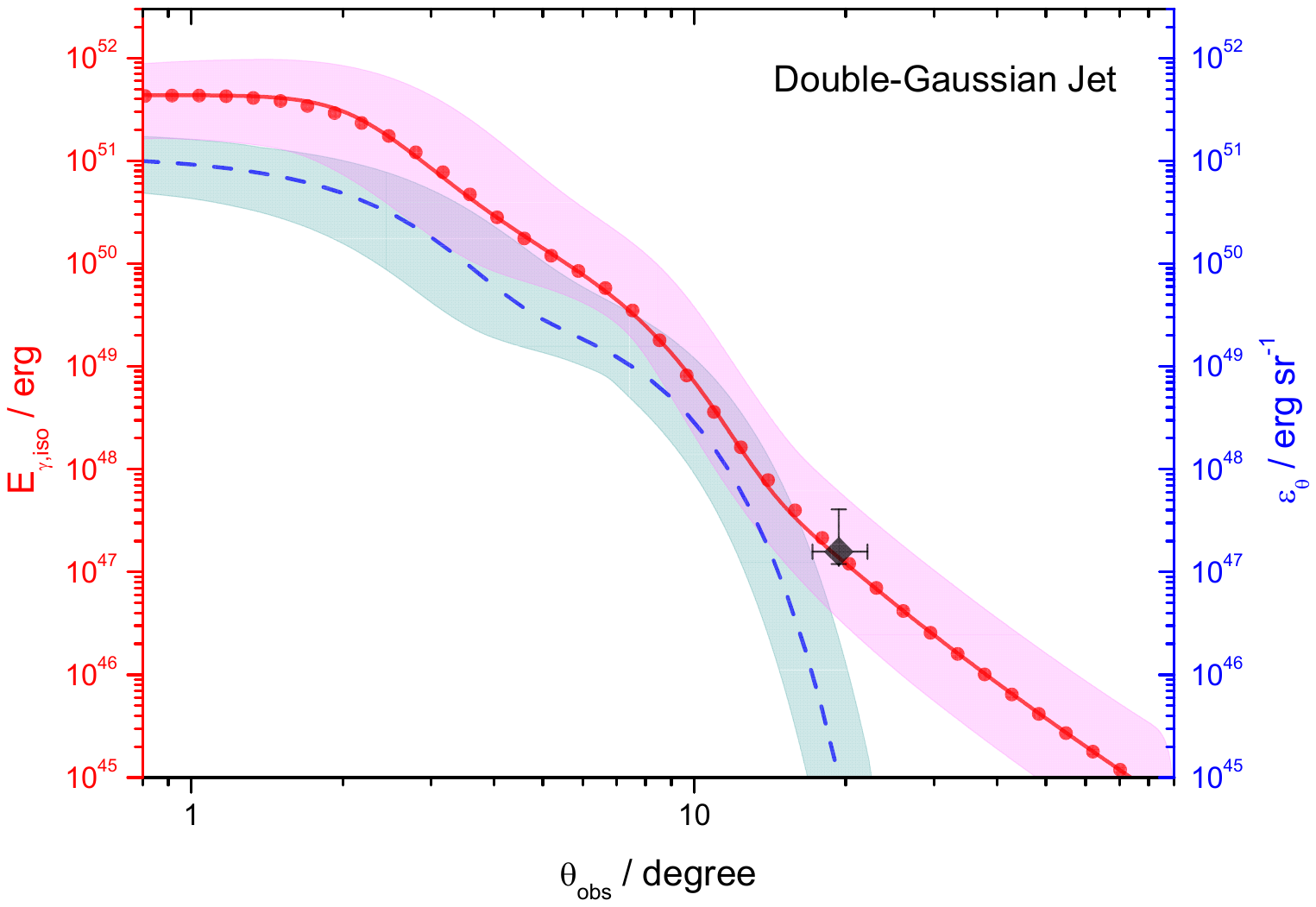}\\
	\caption{The isotropic emission energy ($E_{\gamma,\rm iso}$) of SGRBs as a function of viewing angle (solid red line) in comparison with the distribution of the kinetic energy per solid angle ($\varepsilon_\theta$; dashed 
    blue line). The solid circles are obtained from Eq. (\ref{Lisothetav}), while the solid red lines are plotted by timing Eq. (\ref{LisoLon1}) and (\ref{LisoLon2}) to a constant duration of $T=1$ s, which gives a fit for the solid circles. Here, the central value of the model parameters listed in Table \ref{tab1} are taken and the uncertainty arising from the errors of the parameters are represented by the shaded bands. The value of radiation efficiency $\eta_\gamma$ is adopted to make the solid line be in agreement with the observational emission energy of GRB 170817A (black diamond).} 
	\label{Eiso}
\end{figure}

Because of the jet structure, the emission energy inferred from the observations should highly depend on the viewing angle. Furthermore, due to the relativistic beaming of the jet emission, the observed luminosity could not always be determined by the kinetic energy of the jet in the same direction \citep{Salafia2015}. For a viewing angle $\theta_{\rm obs}$, the isotropically-equivalent energy of the GRB prompt emission can be calculated by \citep{Rybicki1979,Salafia2015}
\begin{eqnarray}
	E_{\gamma,\rm iso}(\theta_{\rm obs})={\eta_{\gamma}}\int {\varepsilon_{\theta}\over \Gamma_{\theta}^4[1-\beta_{\theta}\cos\alpha]^3} d\Omega(\theta,\phi),\label{Lisothetav}
\end{eqnarray}
where $\eta_{\gamma}$ is the radiation efficiency which is assumed to be a constant for different time and different directions. With the parameter values listed in Table \ref{tab1}, we plot the dependence of the emission energy on $\theta_{\rm obs}$ in Figure \ref{Eiso} (solid circles). Radiation efficiency $\eta_{\gamma}$ can be determined by tuning the theoretical lines to be consistent with the observational data of GRB 170817A. For the central value of the parameters, we have $\eta_{\gamma}=0.08$, $0.013$, and $0.2$ for the power-law, single-Gaussian, and double-Gaussian jet structures, respectively. The errors of the structure parameters would lead to the value of $\eta_{\gamma}$ vary within a wide range, which is about an order of magnitude as indicated by the shaded bands in Figure \ref{Eiso}.  

Generally speaking, the emission energy can roughly trace the kinetic energy for viewing angles not much larger than the jet opening angle. However, when the observational direction is far away from the jet axis, the emission energy could become much higher than the kinetic energy at the same direction, particularly, in the case of a Gaussian structure. The large angle emission is actually contributed from the jet material at smaller angles and the prompt emission of GRB 170817A is just in such a situation\footnote{Since the single Gaussian cannot directly explain the prompt emission of GRB 170817A, \cite{Tan2020} suggested an outer Gaussian component. However, as shown here, the GRB 170817A emission actually could be explained by the large-angle emission effect, which indicates that the outer Gaussian may not be indispensable. However, in this paper, we keep considering the double-Gaussian situation, because the double-Gaussian structure could still be a natural result of the jet propagation and it is also helpful for improving the afterglow modeling \citep{Wei2022}.}. 

\begin{table*}
	\begin{center}
		\centering
		\caption{Parameters for the empirical description of the $E_{\gamma,\rm iso}-\theta_{\rm obs}$ relation }\label{tab2}
		\begin{tabular}{cccccccccc}
			\hline\hline
			Model &  $L_{\rm on,GRB 170817A}$ & $\theta_{b1}$ & $\theta_{b2}$ & $\theta_{b3}$ & $s$ & $\alpha_1$ & $\alpha_2$ & $\alpha_3$ & $\mathcal R$\\
			\hline
			Power law  & 51 & 2.5 & 18 & /& 0.6 & 0. & 4.3 & 1.45 & / \\
			Single Gaussian  & 51.5 & 6.0 & 14.3 & / & 0.4 & 0.4 & 9.5 & 5.9 & /\\
			Double Gaussian & 51.62 & 2.1 & 22.5 & 3.65 & 1.4 & 0.0 & 4.9 & 1.75&  0.048 \\
			
			\hline\hline
		\end{tabular}
	\end{center}
\end{table*}

Furthermore, we derive the observational isotropic luminosity from the emission energy by using $L_{\rm iso}=E_{\gamma,\rm iso}/ T$, where $T$ is the duration of the emission. Here, for simplicity, we assume a constant value of $T\sim1$ s for all SGRBs. Therefore, the direction-dependence of the isotropic luminosity can be described analytically by the following functions, which are obtained by fitting the $E_{\gamma,\rm iso}-\theta_{\rm obs}$ relations displayed in Figure \ref{Eiso}. Specifically, for a constant duration, we have
\begin{eqnarray}
	L_{\rm iso}(\theta_{\rm obs})=L_{\rm on} {1 + \left({\theta_{\rm obs}\over \theta_{b2}}\right)^{\alpha_3}\over \left[\left({\theta_{\rm obs}\over \theta_{b1}}\right)^{-\alpha_1s} + \left({\theta_{\rm obs}\over \theta_{b1}}\right)^{\alpha_2s}\right]^{1/s}} \label{LisoLon1}
\end{eqnarray}
for the power-law and single-Gaussian jets and
\begin{eqnarray}
L_{\rm iso}(\theta_{\rm obs})&=&L_{\rm on}\left\{ {1 + \left({\theta_{\rm obs}\over \theta_{b2}}\right)^{\alpha_3}\over \left[\left({\theta_{\rm obs}\over \theta_{b1}}\right)^{-\alpha_1s} + \left({\theta_{\rm obs}\over \theta_{b1}}\right)^{\alpha_2s}\right]^{1/s}}\right.\nonumber\\
&&\left.+\mathcal R \exp\left(-{\theta_{\rm obs}^2\over 2\theta_{\rm b3}^2}\right)  \right\} \nonumber\\ \label{LisoLon2}
\end{eqnarray}
for the double-Gaussian jets, where $L_{\rm on}$ is the luminosity value for the on-axis (i.e., $\theta_{\rm obs}=0^{\circ}$) observation. The invoking of these two empirical functions can help us accelerate the following calculations. It should be kept in mind that the duration of SGRBs is intrinsically determined by the period of the engine activity and, furthermore, influenced by the propagation of the jet through the progenitor material. Nevertheless, ignoring these details, the ultimate distribution of the duration could be well described by a Gaussian as observed, which could not be significantly influenced by the selection effects. Compared with our constant duration assumption, the realistic distribution of the duration would lead to an extra diffuse of the values of the model parameter constrained later in this paper.

As found in \cite{Salafia2015} and \cite{Tan2020}, the $\theta_{\rm obs}$-dependence of the GRB luminosity can substantially influence the determination of the LF of SGRBs, which was usually found to have an apparent broken-power-law form \citep{Wanderman2015,Ghirlanda2016,Tan2020}, where the flat low-luminosity component within the luminosity range ($L_{\rm iso}\sim10^{50}-10^{52}\rm erg~s^{-1}$ can just result from the angular structure of the jet. Therefore, following \cite{Tan2020}, the intrinsic LF of SGRBs which describes the probability distribution of the on-axis luminosity $L_{\rm on}$ of different SGRB jets is assumed to have a simple power law form as
\begin{eqnarray}
	\Phi(L_{\rm on})=\Phi_{*}\left({L_{\rm on}\over L_{\rm on}^{*}}\right)^{-\gamma}\exp\left(-{L_{\rm on}^{*}\over L_{\rm on}}\right),\label{SPL}
\end{eqnarray}
where $\Phi_{*}$ is the normalization coefficient. Then, for an apparent isotropic luminosity $L_{\rm iso}$, its occurring probability should be calculated by integrating over the arbitrary observational directions as
\begin{eqnarray}
	p(L_{\rm iso})=\int\Phi(L_{\rm on}|L_{\rm iso},\theta_{\rm obs}){\sin\theta_{\rm obs}d\theta_{\rm obs}},\label{probability}
\end{eqnarray}
where the fact that the jets can be paired is considered. It is assumed that all SGRB jets have the angular profile identical to that of GRB 170817A, but the total energy of the jets and thus the on-axis luminosity can still be different from each other. In the above integration, the value of $\Phi(L_{\rm on}|L_{\rm iso},\theta_{\rm obs})$ can be obtained by using Eq. (\ref{LisoLon1}) or (\ref{LisoLon2}) to calculate a value of $L_{\rm on}$ corresponding to the given $L_{\rm iso}$ and $\theta_{\rm obs}$ and, subsequently, substitute the value of $L_{\rm on}$ into the expression of $\Phi(L_{\rm on})$.

\subsection{Modeling the flux and redshift distributions of SGRBs}

\begin{figure*}
	\centering\resizebox{0.45\textwidth}{!} {\includegraphics{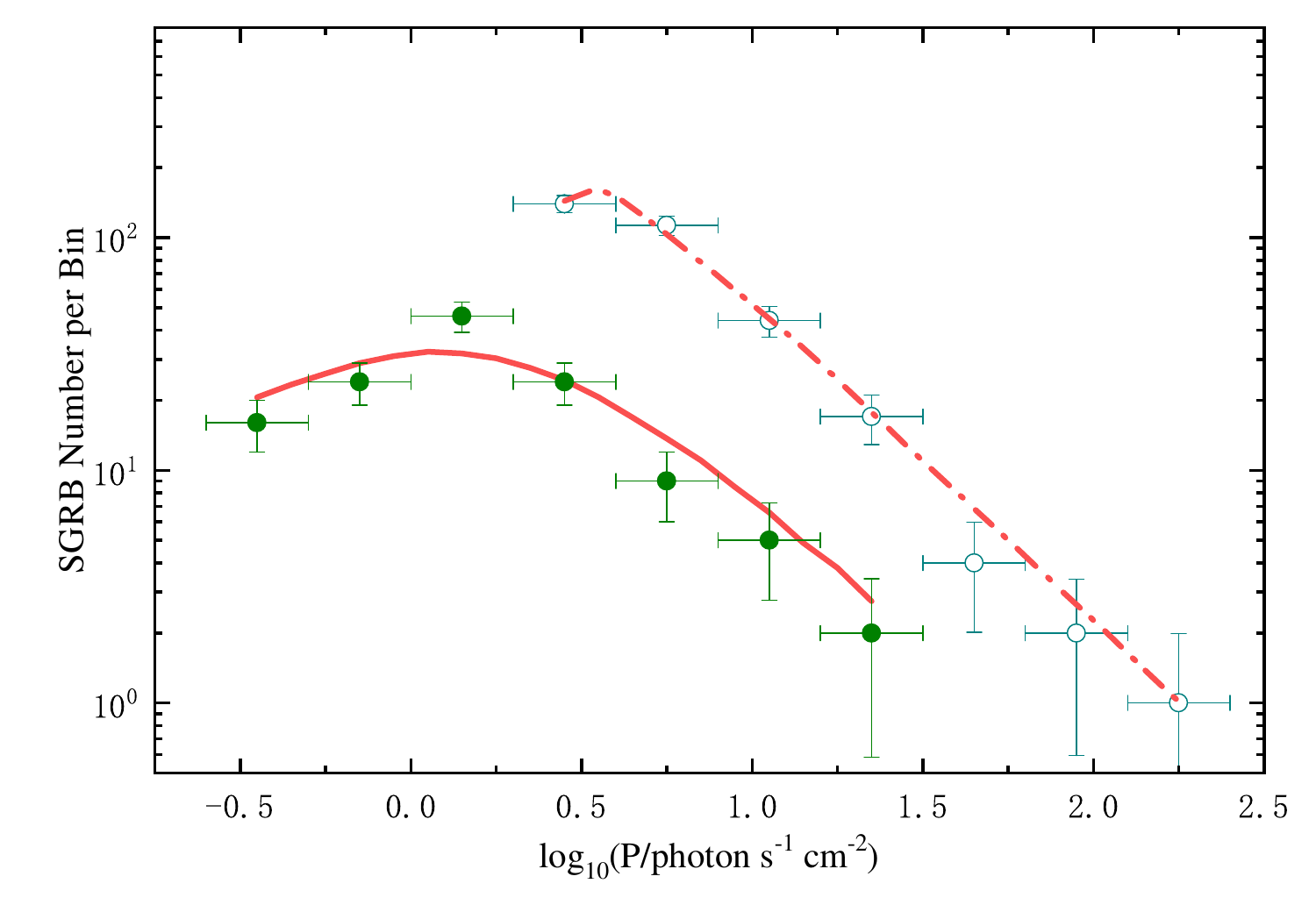}}\resizebox{0.45\textwidth}{!} {\includegraphics{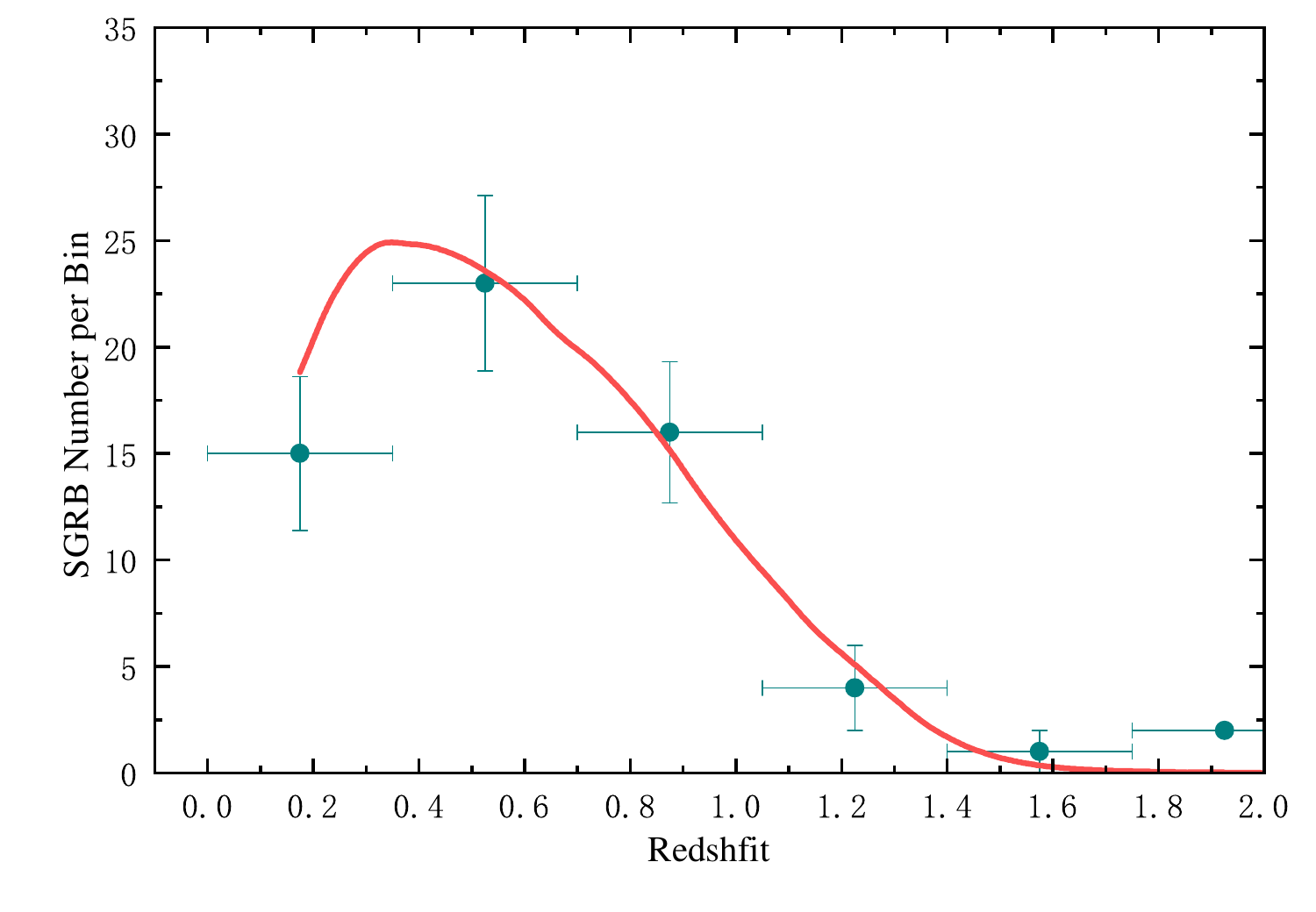}}
	\caption{Comparison between the model-predicted and observational flux and redshift distributions of SGRBs for a power-law jet structure. The solid and open data circles are the results of the Swift and Fermi observations, respectively. 
 }\label{FigPL}
\end{figure*}
\begin{figure*}
	\centering\resizebox{0.45\textwidth}{!} {\includegraphics{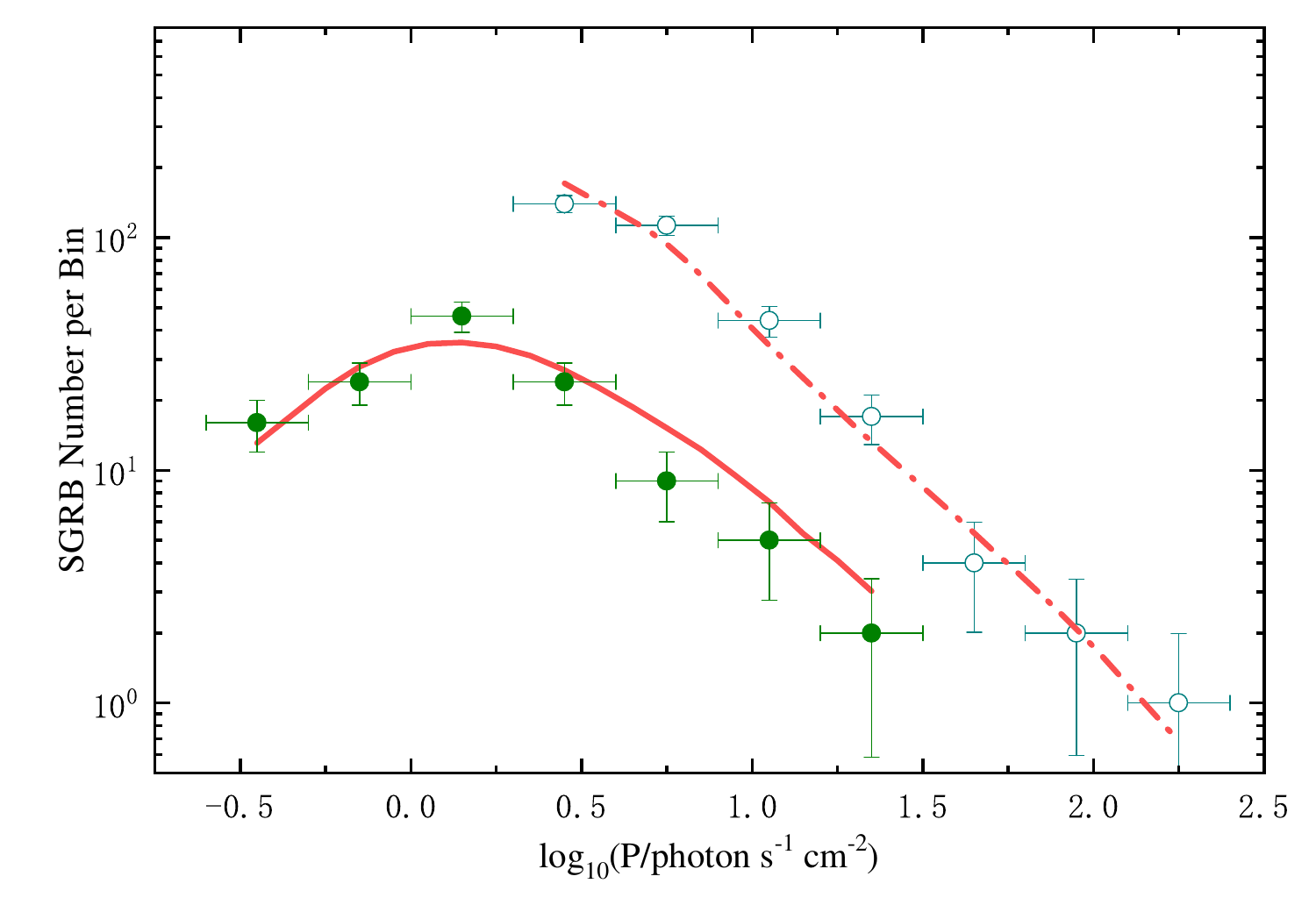}}\resizebox{0.45\textwidth}{!} {\includegraphics{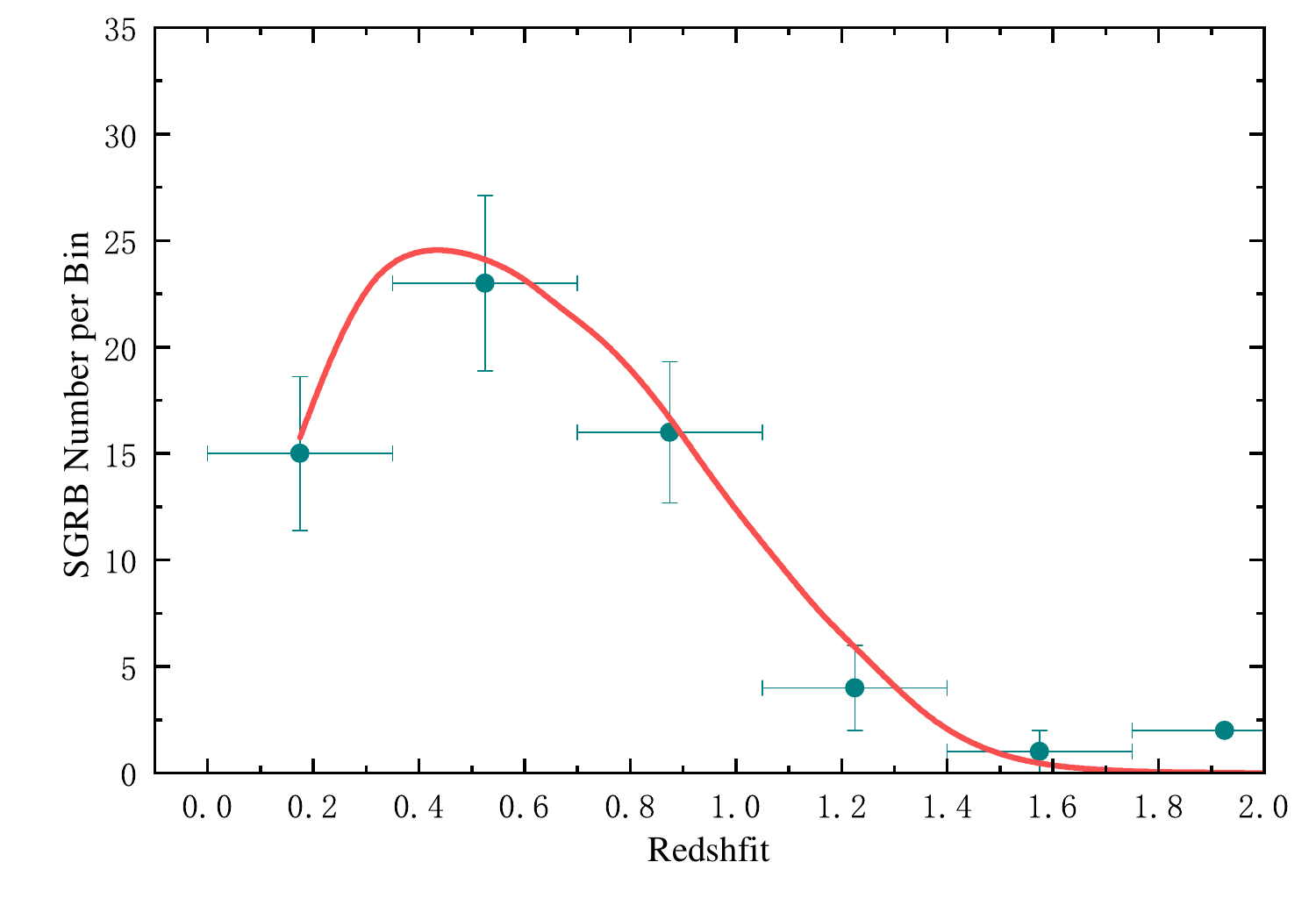}}
	\caption{Same to Figure \ref{FigPL} but for a single-Gaussian jet structure.}\label{FigSG}
\end{figure*}
\begin{figure*}
	\centering\resizebox{0.45\textwidth}{!} {\includegraphics{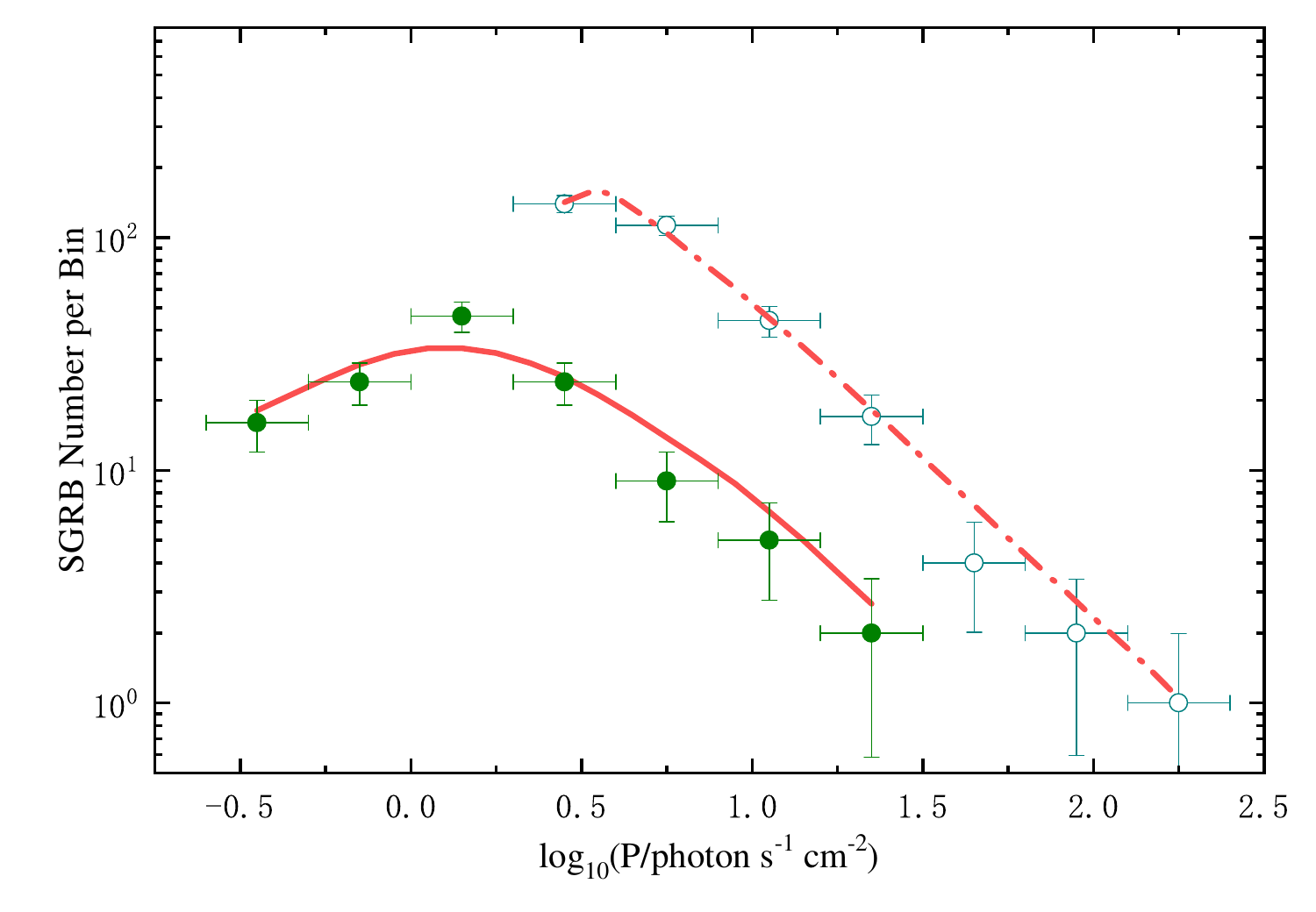}}\resizebox{0.45\textwidth}{!} {\includegraphics{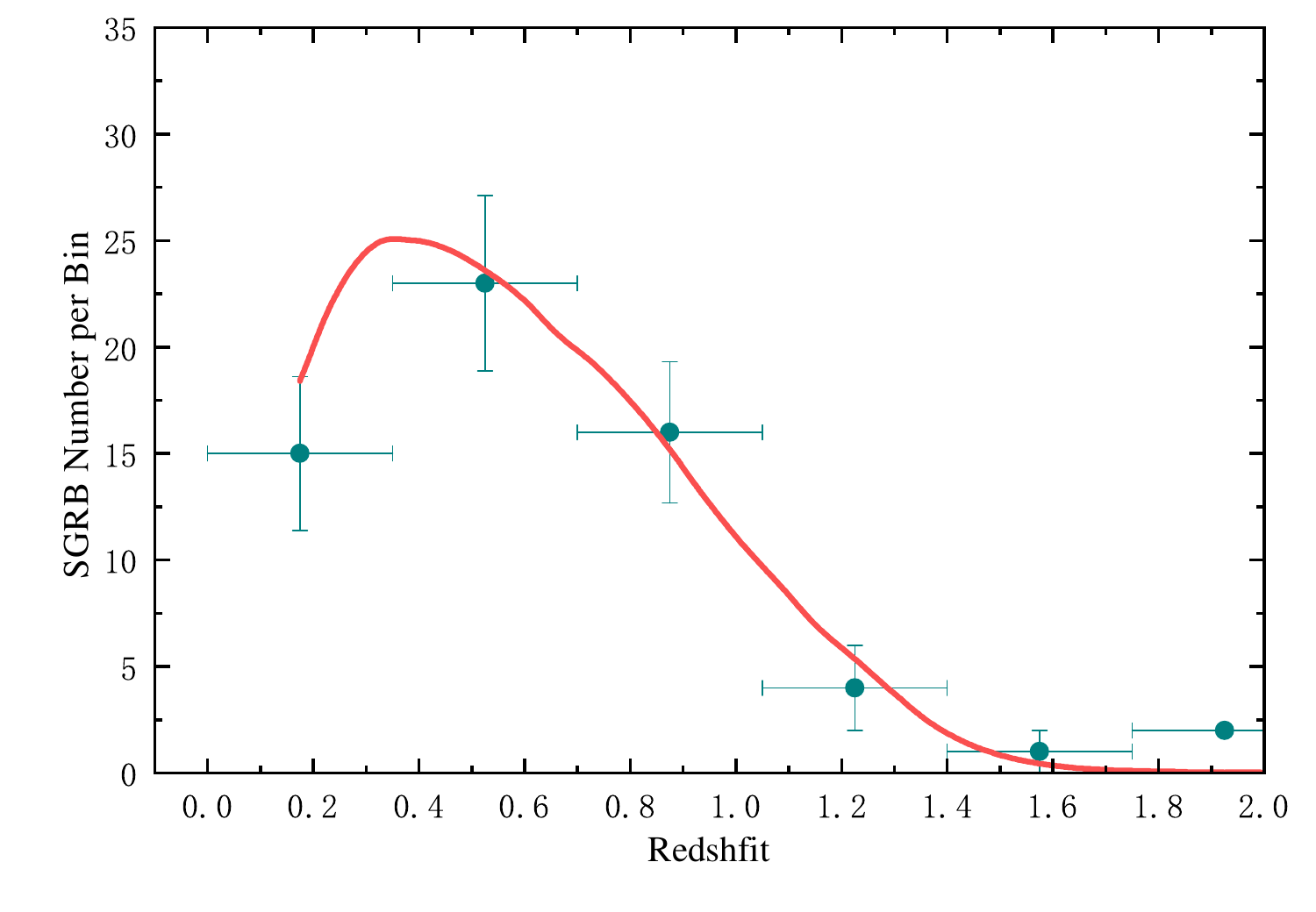}}
	\caption{Same to Figure \ref{FigPL} but for a double-Gaussian jet structure.}\label{FigTG}
\end{figure*}

\begin{table*}
	\begin{center}
		\centering
		\setlength{\tabcolsep}{3.5mm}{}
		\renewcommand\arraystretch{1.4}
		\caption{Constraints on the luminosity function and event rates of SGRBs}\label{tab3}
		\begin{tabular}{cccccccc}
			\hline\hline
			Parameters &  Priors  &&Posteriors&\\
            && Power law  & Single Gaussian  &  Double Gaussian   \\
            \hline
            $\gamma $&(1, 3)& $2.37^{+0.73}_{-0.64}$&    $2.41^{+1.27}_{-0.65}$& $2.25^{+1.66}_{-0.42}$\\
            $L_{\rm on}^{*}/ 10^{52}\erg\s^{-1}$ &(0.1, 10)  & $2.08^{+2.14}_{-1.22}$& $2.62^{+2.23}_{-1.36}$ &$0.56^{+0.68}_{-0.33}$ \\
            $\tau_{\rm min} /\rm Gyr$ & (1, 5) &$3.02^{+0.44}_{-0.55}$ &$3.01^{+0.44}_{-0.55}$ &$2.94^{+0.44}_{-0.55}$ \\
            $\dot{R}_{\rm SGRB}(0)/\rm Gpc^{-3}yr^{-1}$ &  --  &  $325.96^{+39.46}_{-20.59}$&$112.07^{+12.98}_{-8.26}$&	$454.01^{+50.18}_{-33.46}$ \\
            \hline
            $P_{{\rm KS},F}^{\rm Fermi} $& -- & 0.23& $2.76\times 10^{-4}$& 0.20 \\
            $P_{{\rm KS},F}^{\rm Swift}$&-- & 0.13& $0.14$& 0.30 \\
            $P_{{\rm KS},z} $ &-- &0.78 & 0.62 & 0.83\\
         
			\hline\hline
		\end{tabular}
	\end{center}
\end{table*}

For a comparison with the the observational distributions of SGRBs on their fluxes and redshifts, we calculate the model-predicted SGRB numbers in different flux ($P$) ranges and different redshift ($z$) ranges by the following integrations \citep{Tan2020}
\begin{eqnarray}
	N(P_1, P_2)&=&{\Delta\Omega\over 4\pi} \mathcal T \int^{z_{\max}}_{0}\int^{P_{2}}_{P_{1}}\eta(P)\nonumber\\
	&\times&\dot{R}_{\rm SGRB }(z) p(P) dP{dV(z)/dz\over 1+z} dz,
	\label{EQN: PPFD}
\end{eqnarray}
and
\begin{eqnarray}
	N(z_1, z_2)&=&{\Delta\Omega\over 4\pi} \mathcal T \int^{z_2}_{z_1}\int^{P_{\max}}_{0}\eta(P)\vartheta_z(z,P) \nonumber\\
	& \times&\dot{R}_{\rm SGRB }(z)p(P)dP{dV(z)/dz\over1+z}dz, \label{EQN: RD}
\end{eqnarray}
respectively, where $\Delta \Omega$ is the field of view of a telescope, $\mathcal T$ is the working time (a duty cycle of $\sim$50\% for Fermi should be considered \citep{Zhang2018}), $\eta(P)$ and $\vartheta(z,P)$ are the trigger efficiency and the probability of redshift measurement, respectively. In our calculations, the trigger efficiency is considered to increase gradually with the increasing flux, as described by \cite{Howell2014} for the Swift BAT, which reads
\begin{equation}
    \eta(P)={c_1(c_2+c_3P/P_{P}^{*})\over1+P/(c_4P_{P}^{*})},
\end{equation}
where $c_1=0.47, c_2=-0.05, c_3=1.46, c_4=1.45$ and $P_{P}^{*}=1.6\times10^{-7}\rm erg~s^{-1}cm^{-2}$. However, for the Fermi GBM, such an empirical description is absent and thus, we have to choose a sufficiently high hard threshold as $P_{\rm th}=2.37 \rm ph~s^{-1}cm^{-2}$ \citep[e.g.,][]{Wanderman2015}. Correspondingly, the low-flux Fermi data is abandoned. Meanwhile, the redshift selection of the Swift BAT can be expressed as \citep{Cao2011,Tan2020}
\begin{eqnarray}
    \vartheta(z,P)=\min\left[c_5+{P\over P_{z}^{*}},1\right]\exp\left(c_6-{z\over c_7}\right)\left\{1-c_8\exp\left[-{(z-\mu)^2\over 2\sigma^2}\right]\right\}
\end{eqnarray}
where $c_5=0.27, c_6=0.3, c_7=8.9, c_8=0.41, P_{z}^{*}=2.0\times10^{-6}{\rm erg~s^{-1}cm^{-2}}, \mu=1.60$, and $\sigma=0.23$. The occurring probability $p(P)$ of SGRBs can be obtained by substituting the corresponding value of $L_{\rm iso}$ into Eq. (\ref{probability}). The relationship between luminosity and flux is given by $L_{\rm iso}=4\pi d_{\rm L}^2Pk$, where $d_{\rm L}$ is the luminosity distance and $k$ is a correction factor that converts the observational photon flux in the detector band to the energy flux in a fixed rest-frame band of $(1,10^4)$ keV. More specifically, the calculation of the $k-$correction involves a tentative value of the peak energy of the Band spectrum ($E_{\rm p}$) of SGRBs, which can be derived from an empirical $E_{\rm p}-L_{\rm on}$ correlation as presented in Eq. (7) in \cite{Tan2020}. The comoving cosmological volume element in the integration is defined as $dV(z)/dz=4\pi d_{\rm c}(z)^2c/H(z)$, where $d_{\rm c}(z)=c\int_0^zH(z')^{-1}dz'$, $H(z)=H_0[(1+z)^3\Omega_{\rm m}+\Omega_{\Lambda}]^{1/2}$ and the
cosmological parameters are taken as $\Omega_{\rm m}=0.315$,
$\Omega_\Lambda=0.685$, and $H_0=67.4\rm km~s^{-1}Mpc^{-1}$ \citep{Planck2020}. 
Finally, the limit values of the redshift $z_{\max}$ and the flux $P_{\max}$ are taken according to the boundaries of the observational ranges.

The most crucial input of the above integrations is the cosmic rate of SGRBs, which can usually be connected with the cosmic star formation rates (CSFRs) by a delay time since the SGRBs are produced by mergers of compact binaries, where the delay time is determined by the formation process of the
compact binaries and the orbital decay through gravitational radiation. By assuming the delay time $\tau$ distributes with a probability function $F(\tau)$, we can express the cosmic SGRB rates by \citep[e.g., ][]{Regimbau2009,Zhu2013,Regimbau2015}:
\begin{eqnarray}
	\dot{R}_{\rm{SGRB}}(z)&\propto&(1+z)\int_{\tau_{\min}}^{t(z)-t(z_{\rm b})}{\dot{\rho}_{*}[t(z)-\tau]\over 1+z[t(z)-\tau]}F(\tau)d\tau\nonumber\\
	&\propto&(1+z)\int^{z_{\rm b}}_{z[t(z)-\tau_{\min}]}{\dot{\rho}_{*}(z')\over 1+z'}F[t(z)-t(z')]{dt\over dz'}dz'\label{sgrbr},
\end{eqnarray}
where $\dot{\rho}_{*}(z)$ is the CSFR,  $t(z)=\int_z^{\infty}[(1+z')H(z')]^{-1}dz'$, $dt/dz=-[(1+z)H(z)]^{-1}$, and $z_{\rm b}$ represents the redshift at which the binaries started to be formed. Finally, please see \cite{Cao2011} and \cite{Tan2020} for the details of the expressions of $k$, $\dot{\rho}_{*}(z)$, and $F(\tau)$.

\begin{figure}
	\centering\resizebox{0.45\textwidth}{!}
	{\includegraphics{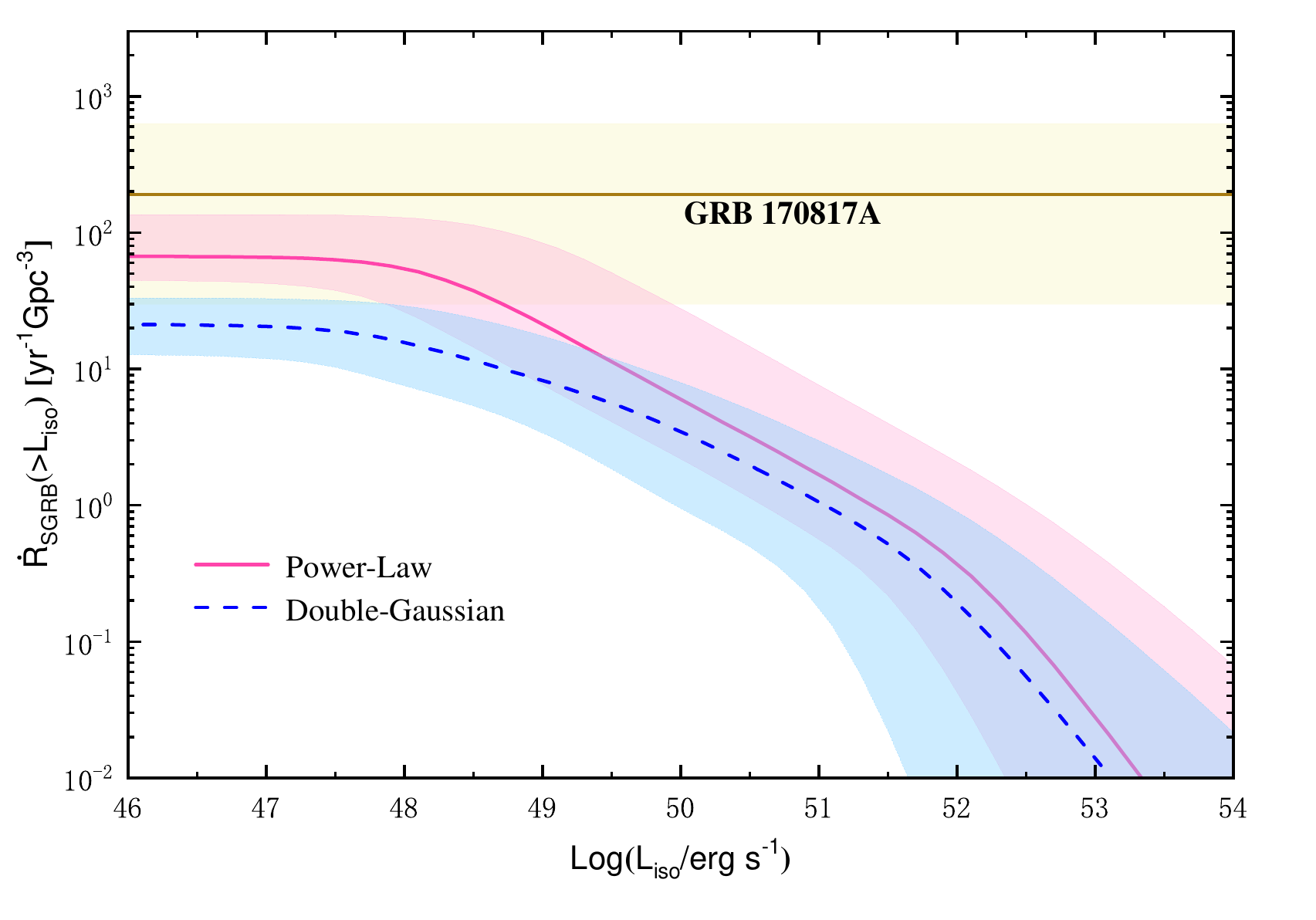}}
	\caption{The apparent LF given by the power-law and the double-Gaussian jet models, in comparison with the event rate $\rm 190^{+440}_{-160} yr^{-1}Gpc^{-3} $ inferred from the GRB 170817A observations represented by the shaded band \citep{Zhang2018}.}
	\label{fig6}
\end{figure}

Assigning values for parameters $\gamma$, $L_{\rm on}^{*}$, and $\tau$, we generate a number of $10^5$ simulated SGRB samples according to the occurring probability given by Eqs. (\ref{probability}) and (\ref{sgrbr}). Then, we can obtain the number distributions of these mocked samples on their redshifts or fluxes and, furthermore, compare these distributions with the observational ones. Finally, the value of $\dot{R}_{\rm SGRB}(0)$ is determined by normalizing the mocked distributions by the total number of Swift SGRBs. Here, for the Swift SGRBs, we extract their 1-second peak photon fluxes in the $15-150\kev$ energy range from \href{https://swift.gsfc.nasa.gov/results/batgrbcat/index.html}{https://swift.gsfc.nasa.gov/results/batgrbcat/index.html}, up until 2023. Meanwhile, for the Fermi SGRBs, their 64 ms peak photon fluxes in the energy band of $50-300 \kev$ have been provided at \href{https://heasarc.gsfc.nasa.gov/W3Browse/fermi/fermigbrst.html}{https://heasarc.gsfc.nasa.gov/W3Browse/fermi/fermigbrst.html}, where the data is updated until 2018. In principle, these peak fluxes can be somewhat different from the average ones we used in the model, which is nevertheless not very significant for a short period of $T\sim 1$s for the average ones. Furthermore, considering that the difference between the peak and average fluxes can be actually coupled with and then included into the assumed constant value of $T$, we ignored this difference in our calculations for simplicity. The best fit of the observational distributions, which is presented in in Figures \ref{FigPL}, \ref{FigSG}, and \ref{FigTG} for the three jet structures, is obtained by minimizing the sum of $\chi^2$ of the redshift and flux distributions. Here, we calculate $\chi^2 =\sum_{i=1}^{n} \frac{(O_i - M_i)^2}{M_i} $ with \( O_i \) and \( M_i \) being the observational and simulated SGRB numbers, respectively, in the corresponding redshift or flux bins. The priors and posteriors of our simulations are listed in Table \ref{tab3}. Finally, as indicated by the KS tests of the best fits, the single-Gaussian model could be relatively disfavored by the peak flux distribution for the Fermi data, even though the large-angle emission effect is taken into account. In comparison, the goodness of the power-law and double-Gaussian models is basically comparable to each other (the latter is relatively better). With the obtained values of the intrinsic LF parameters, the typical delay time, and the total rates of SGRBs, we plot the luminosity dependence of the event rates of the SGRBs (e.g., the apparent LF) in Figure \ref{fig6}, compared to the event rate of $\rm 190^{+440}_{-160} yr^{-1}Gpc^{-3}$ inferred by GRB 170817A \citep{Zhang2018}. As shown, the power-law structure can determine an event rate for $L_{\rm iso}\gtrsim 10^{47}\rm erg~s^{-1}$ well consistent with the central value of the GRB 170817A rate, whereas the double-Gaussian model can only reach the marginal value of the error range.

\section{Summary}
The uncovering of the jet structure is one of the most crucial aims of the GRB research, which can help to understand the nature of their progenitors and central engines. It has been long expected to constrain the jet structure by using the observed afterglow emission. However, usually, only an effective half-opening angle of the jet can be derived from the data, if a so-called jet break characteristic can be identified from the afterglow light curves. The discovery of GRB 170817A had changed this awkward situation, because it was significantly observed off-axis. Then, an immediate question is that whether the jet structures inferred from the afterglows of GRB 170817A can be compatible with the statistical properties of the SGRB population, if it is assumed that all SGRBs including GRB 170817A have a common origin and explosive mechanism. Therefore, in this paper, we investigate three typical empirical jet structures including the power-law, single-Gaussian, and double-Gaussian cases, the parameters of which are first constrained according to the afterglow data of GRB 170817A. It is further demonstrated that these three types of jet structure can all account for the prompt luminosity of GRB 170817A, with the consideration of the large-angle emission effect of relativistic jets. However, the single-Gaussian structure could somewhat fail to reproduce the flux distribution for the Fermi data. By comparison, the power-law structure is more favored by the statistical results and, furthermore, predicts an event rate closer to the value inferred from GRB 170817A. Nevertheless, it should be kept in mind that the large number of model parameters could still make these conclusions subject to the large uncertainty of the parameters.

\begin{acknowledgements}
This work is supported by the National Key R\&D Program of China (2021YFA0718500),
the China Manned Spaced Project (CMS-CSST-2021-A12),
the National Natural Science Foundation of China (grant Nos. 12393811, U1838203, and 11803007), and Hubei Provincial Outstanding Young and Middle-aged Science and Technology Innovation Team Project of China (T2021026).
\end{acknowledgements}


\bibliographystyle{raa}

\label{lastpage}

\end{document}